# ENERGY CONVERSION USING NEW THERMOELECTRIC GENERATOR

*G. Savelli[1,2], M. Plissonnier[1], J. Bablet[1], C. Salvi[1], J.M. Fournier[1,2]*

[1] CEA/LITEN/DTNM/LCH, 17 rue des Martyrs, 38054 Grenoble Cedex 9, France
[2] LEG, CNRS UMR 552, BP 46, 38402 St Martin d'Hères Cedex, France

## ABSTRACT

During recent years, microelectronics helped to develop complex and varied technologies. It appears that many of these technologies can be applied successfully to realize Seebeck micro generators: photolithography and deposition methods allow to elaborate thin thermoelectric structures at the micro-scale level.

Our goal is to scavenge energy by developing a miniature power source for operating electronic components. First Bi and Sb micro-devices on silicon glass substrate have been manufactured with an area of $1cm^2$ including more than one hundred junctions. Each step of process fabrication has been optimized: photolithography, deposition process, anneals conditions and metallic connections. Different device structures have been realized with different micro-line dimensions. Each devices performance will be reviewed and discussed in function of their design structure.

## 1. INTRODUCTION

Since the last decade, there is a growing interest of wireless sensor nodes with goals of monitoring human environment. Because advances in low power VLSI design and CMOS fabrication have dramatically decreased power requirements of sensors, it is now possible to consider self-powered system in sensor node [1].

In the same time, the interest in producing micro-electromechanical systems (MEMS) opens new opportunity in the field of micro power generation. Micro thermoelectric converters are a promising technology due to the high reliability, quiet operation and are usually environmentally friendly. The efficiency of a thermoelectric device is generally limited to its associated Carnot Cycle efficiency reduced by a factor which is dependent upon the thermoelectric figure of merit (ZT) of the materials [2] used in fabrication of the thermoelectric device. Recent developments in micro-thermoelectric devices using thin film deposition technology [3] have shown that energy scavenged from human environment (low thermal gradient) matches energy sensor's power need.

The used thermoelectric materials (TE) are here bismuth and antimony. Both Bi and Sb are semimetals, that is, there is an energy overlap between the valence and conduction bands. Near room temperature, the thermoelectric power (or Seebeck coefficient) of both Bi and Sb are enough small: typical values are about $-70\mu V.K^{-1}$ for Bi and $40\mu V.K^{-1}$ for Sb [4]. Nevertheless, the acquired experience with these usual thermoelectric materials for our devices will have been very enriching and helpful for our presently device improvement with more competitive thermoelectric materials.

Thus, in this paper we review a wafer technology approach to manufacture a thermoelectric device. Design and technological process steps will be identified in order to propose a strategy to manufacture thermoelectric converters (TEC).

## 2. EXPERIMENTAL

### 2.1. Device description

We use a four inches glass substrate. 42 chips and 6 test areas are distributed on it. Fig.1. shows this Bi-Sb prototype.

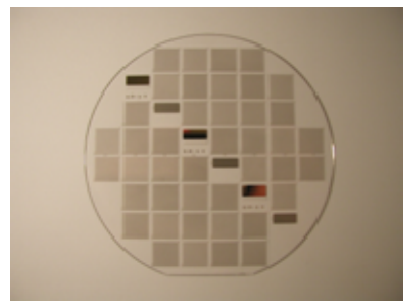

Fig.1. Bi-Sb device realized by microelectronic technologies: PVD deposition, photolithography….

Dimension of chips are $1x1cm^2$, bismuth and antimony lines widths are 20, 30 or 40µm, spaced by 20µm.
This geometry is obtained by photolithography with a serial of 3 masks.

These geometries allow us to obtain 250 lines (125 in bismuth and 125 in antimony in alternation), i.e 125 junctions for the 20x20 chips, 208 lines and 104 junctions for the 30x20 chips and 166 lines and 83 junctions for the 40x30 chips.

Moreover Bi and Sb lines are electrically connected in series by using Ti and Au metallic junctions. Fig.2. shows these lines more in details and an enlargement of the metallic connections.

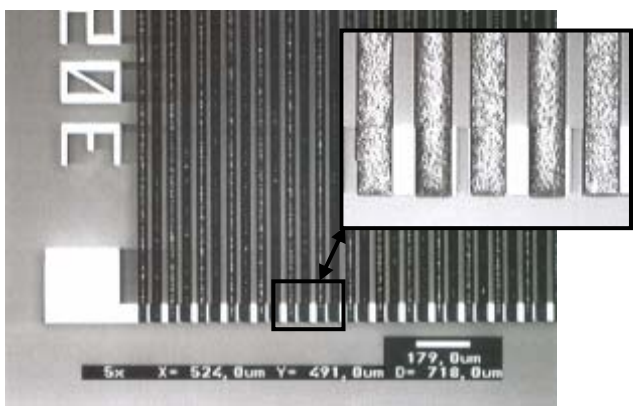

Fig.2. Bismuth and antimony lines with electric connections.

Bismuth and antimony lines, and titanium and gold connections are deposited by sputtering PVD (Physical Vapor Deposition) system. This deposition choice is explained in the next paragraph. Sputtering process use six inches targets for a best thickness uniformity. Each deposition is realized in Ar atmosphere at 1.2 Pa. In addition, an rf generator provides power supplies from 0 to 300W and operates at a frequency of 13.56MHz.

Also it is necessary to add adhesion layers for bismuth and antimony deposition. These sublayers permit Bi and Sb to adhere to the substrate during photolithography operations (resin addition, etching…). A flash Ti adhesion layer (1nm) is used for Bi lines and flash Ti+Au adhesion layers (1nm+1nm) are used for Sb lines. We will check in part 3. that these added layers have no influence on both thermoelectric results and crystallographic structures.

**2.2. Device treatment and characterization**

Annealing is the crucial point for this device. Indeed in thermoelectrics, it is important to obtain the lowest electrical resistivity. To succeed in these lowest values high layers quality are necessary. But, for example, bismuth is well known to deposite in the form of grains [5-6] which increases resistivity. Objective is thus to improve layers quality, and so to reduce grains size. For that, in first, sputtering deposition is used because, in PVD, it allows to obtain grain sizes smaller than with evaporation deposition [5]. Furthermore different annealing conditions have been studied to reduce grain sizes. Two kinds of annealings will be compared in section 3: annealing by furnace and by laser. Annealings by furnace are realized at 260°C for bismuth ($T_m$[Bi-Ti]= $T_m$[Bi]=271°C) and 355°C for antimony ($T_m$[Sb-Au]=360°C) with Ar atmosphere for 8 hours ($T_m$ are obtained according to Binary Alloy Phase Diagrams). Annealing by laser are realized using a Xe-Cl excimer laser with 200ns of impulsion length and a wavelength of 308nm.

To analyse our devices performances, we use several characterization tools: Scanning Electron Microscopy (SEM) to check geometry dimensions and photolithography quality, X-rays diffraction to control crystallographic structure of bismuth and antimony, a four probe method is used to measure electrical resistivity, and a thermoelectric characterization tool which provides us Seebeck coefficient and useful electrical power.

### 3. RESULTS

**3.1. Annealings influence**

In this part, we present results obtained before and after annealing and compare the two methods. Here cristallographic structures and electrical resistivity are studied. Firstly, annealings influence on bismuth lines is tested.

As Bi layers are "granular", resistivity is too high. Thus Bi resistivity before annealing is 1600µΩ.cm.

To decrease this value, we tested two kinds of annealing. Fig.3. shows Bi lines before and after the two kinds of annealing (on this Fig.3. just Bi lines are deposited).

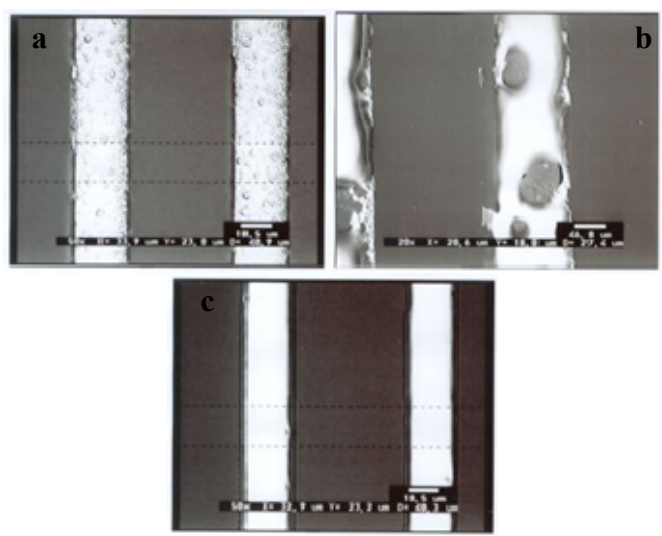

Fig.3. Influence of annealing on Bi lines - a: without annealing - b: annealing by furnace at 260°C - c: annealing by laser.

Fig.3.a. confirms the granular aspect of Bi lines. We see with fig.3.b. that some parts of lines melted and can involve too important discontinuities in lines. Resistivity after annealing by furnace (at 260°C for 8 hours) decreases to 900μΩ.cm.

Fig.3.c. shows Bi lines after annealing by laser. The granular aspect seems to have disappeared. Fig.4. shows the influence of the annealing by laser.

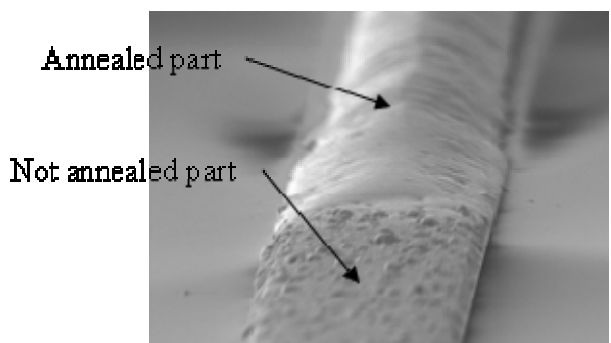

Fig.4. Bi line: one part is annealed by laser, not the other one (photo realized by SEM).

On Fig.4. we chose voluntary to anneal just one part of the line to see the difference. We see that grains disappeared on the annealed part. Fig.5. shows a cross-section of the line (annealed part): absence of grains is confirmed.

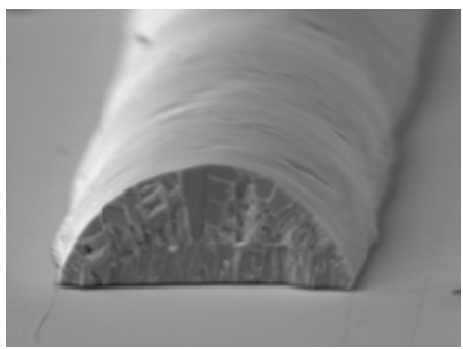

Fig.5. Cross section of the Bi annealed line (photo realized by SEM).

The granular aspect of Bi layer is reduced. This is confirmed by resistivity measurements. We obtain a resistivity of 800μΩ.cm after annealing by laser.

Annealing influence on Sb layers is also reviewed with different annealing conditions. Deposited Sb resistivity is 1100μΩ.cm.

After an annealing of 355°C for 8 hours, its resistivity decreases to 825μΩ.cm. There is no notable improvement after an annealing by laser, its resistivity is 825μΩ.cm too.

Fig.6. shows two lines, one of bismuth and one of antimony, where each line has an annealed part and not the other one.

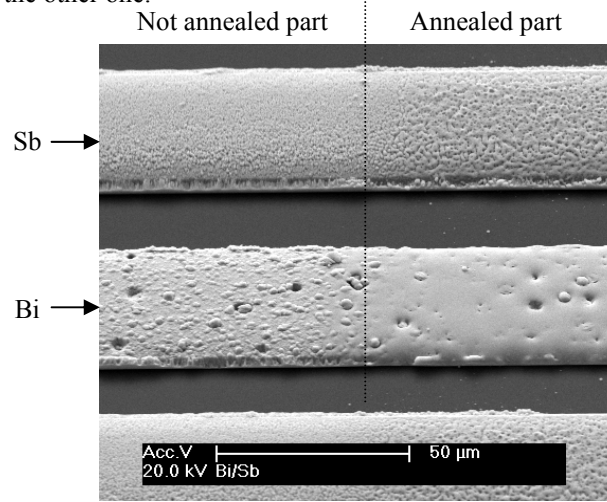

Fig.6. Influence of annealing by laser on both Bi and Sb lines.

We always see that Bi granular aspect disappears and that there is no important visible change concerning Sb layer. To confirm these observations, Fig.7. present bismuth and antimony X-rays diffraction spectra.

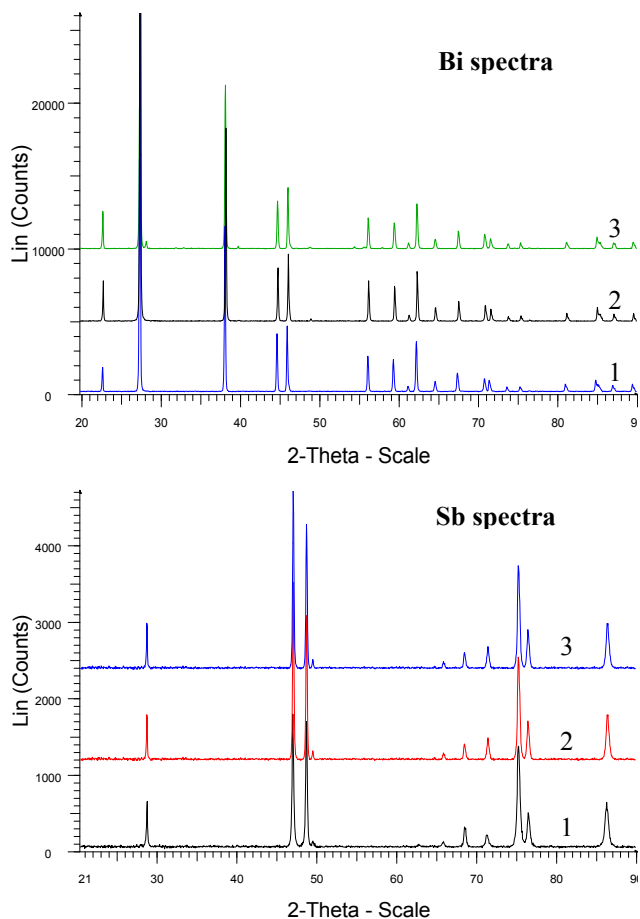

Fig.7. Bi and Sb X-rays spectra - 1: without annealing - 2: after annealing by laser - 3: after annealing by furnace.

First, we can observe that we obtain a good crystallographic structure, i.e. a rhombohedral structure for bismuth films. All peaks are identified by the JCPDS files. These parameters are: a=b=4.547Å, c=11.8616Å, and α=β=90°, γ=120°. For antimony films, we obtain a rhombohedral structure too. In the same way, its structure is identified by the JCPDS files, with the following parameters: a=b=4.307Å, c=11.273, and α=β=90°, γ=120°.

These spectra show that Bi and Sb annealing by laser don't disturb the initial polycrystalline structure, which is an important point, because this structure, which can be considered as a slightly distorted cube, is responsible for a minute band overlap, leading to the presence of small and equal electron and hole densities at all temperatures [7]. Moreover, any peak reveals the presence of titanium or gold, which ensures that any Bi-Ti or Sb-Au alloy appeared during annealing.

To resume, Fig.8. shows a table summarizing resistivity results for Bi and Sb, with or without annealing.

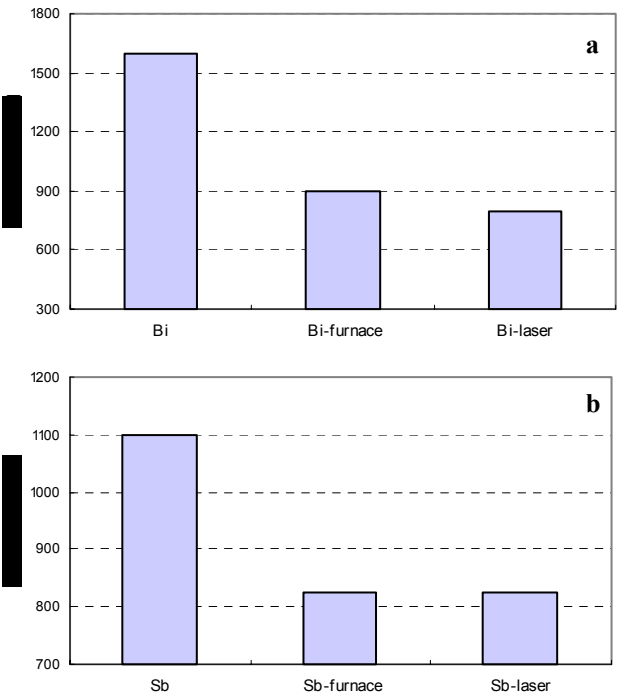

Fig.8. Influence of annealing on electrical resistivity - a: for Bi lines - b: for Sb lines.

Finally, we realize a last annealing on the final device at 240°C for 9 hours in Ar atmosphere. This annealing permits to improve interface layers and to stabilize the device.

### 3.2. Thermoelectric results

Thermoelectric generators are characterized by the Seebeck voltage $V_S$ and the useful electrical power $P_u$, defined by the following equations:

$$P_{cc} = \frac{V_s^2}{R_g} \quad ; \quad P_u = \frac{P_{cc}}{4} \qquad (eq.1)$$

$$R_g = N\left[\rho_{Bi}\frac{L_{Bi}}{A_{Bi}} + \rho_{Sb}\frac{L_{Sb}}{A_{Sb}} + 2\rho_m\frac{L_m}{A_m}\right] \qquad (eq.2)$$

$$R_g \approx N\left[\rho_{Bi}\frac{L_{Bi}}{A_{Bi}} + \rho_{Sb}\frac{L_{Sb}}{A_{Sb}}\right] \quad \text{with } \rho_m \ll \rho_{Bi} \text{ and } \rho_{Sb}$$

where $P_{cc}$ is the short-circuit power, $R_g$ is the chip global electric resistance, N is the total junctions number for one chip, ρ is the electrical resistivity, L is the line length and A the line section area. Metallic - TE materials contact resistance is here neglected in front of global lines resistance.

$V_s$ depends on Seebeck coefficient (or thermoelectric power S) and the number of electrically connected junctions in series. Thus $V_s$ increases when lines density increases. In the same way, useful electrical power $P_u$ depends on $V_s$ and $R_g$ (eq.1). In that case, lines geometry impacts directly to $P_u$. Moreover, eq.2 predicts $R_g$, and so $V_s$ and P. Thus using eq.2, we report on Fig.9. $R_g$ values for each device geometry: $R_{g\ theo.}$ is calculated with bulk material resistivity ($\rho_{Bi}$=117μΩ.cm, $\rho_{Sb}$=40.1μΩ.cm at 300K and $\rho_m$ is the metallic junctions resistivity). $R_{g\ cal.}$ is calculated with experimentally measured Bi-Sb resistivity (i.e. after annealing by laser). And finally, $R_{g\ eff.}$ is the effective device electrical resistance after the last annealing. Moreover, as expected, $R_{g\ eff.}$ depends strongly on lines geometry and lines density.

|  | **20x20** | **30x20** | **40x20** |
|---|---|---|---|
| **N** | 125 | 104 | 83 |
| **$R_{g\ theo.}$ (kΩ)** | 17,8 | 9,9 | 5,9 |
| **$R_{g\ cal.}$ (kΩ)** | 184,7 | 102,4 | 61,3 |
| **$R_{g\ eff.}$ (kΩ)** | 82 | 63,8 | 31 |

Fig.9. Comparison of $R_g$ values using different resistivity values.

Results show that annealing process is critical in order to decrease $R_g$ to the minimal value $R_{g\ theo.}$ Last annealing gave a decrease of 55% of the chip global resistance. Moreover in any case, $R_g$ values decrease normally by increasing lines width and so, decreasing lines number.

Fig.10. shows the useful electric power $P_u$ as a function of temperature gradient and using these different values of $R_g$ for the 40x20 device.

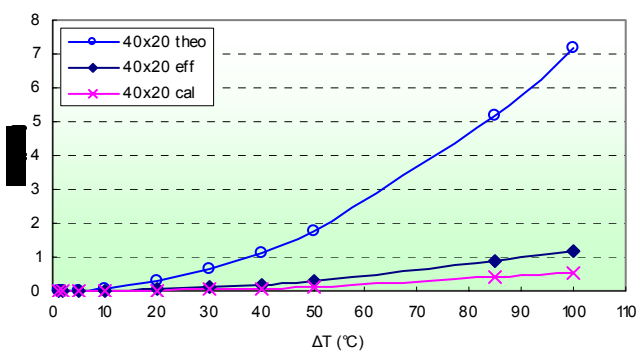

Fig.10. Useful electric power $P_u$ as a function of temperature gradient and $R_g$.

This graph shows that using bulk material resistivity values, we can obtain 7.2μW for ΔT=100K. Without the last annealing we obtain 0.65μW and with this last annealing, we obtain 1.2μW for ΔT=100K, which confirm a decrease of nearly 50%.

It is interesting to compare the Seebeck voltage (Fig.11.a.) and useful electrical power (Fig.11.b.) as a function of lines geometry.

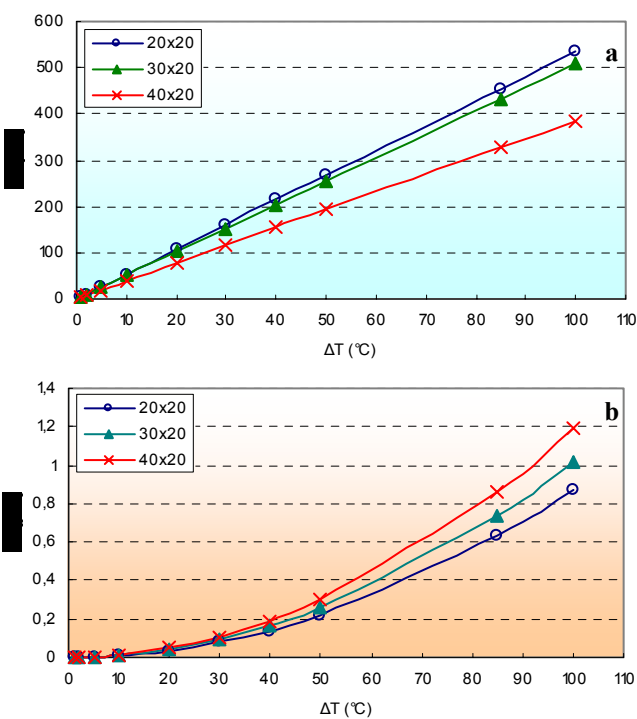

Fig.11. Seebeck voltage $V_s$ evolution (a) and useful electrical power $P_u$ evolution (b) as a function of temperature gradient and geometry.

For the Seebeck voltage, as explained before, we obtain the most important value (535mV for ΔT=100K) for the device which has the highest junctions numbers (i.e. 20x20). On the contrary and as shown in Fig.9., concerning $P_u$, we obtain the highest value (1.2μW for ΔT=100K) for the device which has the lowest junctions numbers (i.e. 40x20). These power values are enough low, mainly due to global electrical resistances, and so due to device geometry.

## 4. CONCLUSION

Design and technological process steps were reviewed and thermoelectric device performances have been discussed. Different annealing conditions demonstrated their strong influence on improving Bi-Sb electrical resistivity and consequently increasing thermoelectric generator performances. Seebeck voltage and electrical power have shown a high voltage and low power density for each device geometry. This means that these configurations are not enough well adapted to supply high power but are interesting for temperature sensors thanks to a good sensitiveness in voltage.

Nevertheless, we currently study 3D device geometry better adapted for higher powers including our process steps optimization performed during this work.

New high performance thermoelectric materials, environmentally friendly as doped silicon and silicon germanium alloys, will be process taking design optimization, process steps and electrical characterization into account.